\documentclass[runningheads]{svmult}

\usepackage{makeidx}   
\usepackage{graphicx}  
\usepackage{subeqnar}  
\usepackage{multicol}  
\usepackage{physprbb}  
\makeindex             
\usepackage{lscape}
\usepackage{float}

\begin{document}
\title*{Optical Observations of the Dark Gamma-Ray Burst GRB 000210}
\toctitle{Optical Observations of the GRB 000210 Error Box}
%
%
\titlerunning{Optical Observations of the GRB 000210 Error Box}
%
\author{Javier  Gorosabel\inst{1}
\and Jens       Hjorth\inst{2}
\and Holger     Pedersen\inst{2}
\and Brian   L. Jensen\inst{2}
\and Lisbeth F. Olsen\inst{2}
\and Lise Christensen\inst{2}
\and Evencio    Mediavilla\inst{3}
\and Rafael     Barrena\inst{3}
\and Johan   U. Fynbo\inst{4}
\and Michael I. Andersen \inst{5}
\and Andreas O. Jaunsen\inst{4}
\and Stephen    Holland\inst{6}
\and Niels      Lund\inst{1}}
\authorrunning{Gorosabel et al.}
%
%
\institute{Danish Space Research Institute, Juliane Maries Vej 30, DK-2100 Copenhagen \O , Denmark
\and Astronomical Observatory, University of Copenhagen, Juliane Maries Vej
30, DK-2100 Copenhagen \O, Denmark
\and Instituto de Astrof\'{\i}sica de Canarias, E-38200 La Laguna,
Tenerife, Canary Islands, Spain
\and European Southern Observatory, Karl-Scwharzschild-Stra\ss e, D-85748 Garching, Germany
\and Division of Astronomy, P.O. Box 3000, FIN-90014 University of Oulu, Finland
\and Department of Physics, University of Notre Dame, Notre Dame, IN
46556-5670, USA
}

\maketitle              

\begin{abstract}
  We report on optical observations on GRB  000210 obtained with the 2.56-m
  Nordic Optical Telescope and  the  1.54-m Danish Telescope starting  12.4
  hours after the  gamma-ray event.   The content of  the X-ray  error  box
  determined by the Chandra satellite is discussed.
\end{abstract}

\section{Introduction}
The BeppoSAX Gamma  Ray Burst Monitor and   the Wide Field  Camera (unit 1)
observed  a  strong  gamma-ray burst  on   2000 February 10.36396   UT.  It
exhibited an X-ray flux of 7.4 Crab (2-26 keV) and a duration of $\sim$ 20s
\cite{Stor00}.  The  field was observed  by   the Chandra  X-ray  satellite
approximately  21  hours  after the   GRB  \cite{Garcia00a,Garcia00b} which
localised  an uncataloged X-ray source to  within 2$^{\prime \prime}$.  The
position  was consistent with the one  derived  independently by BATSE, IPN
and     the     Narrow-Field   Instruments  (NFI)       on-board   BeppoSAX
\cite{Kippen00,Hurley00,Costa00}. The Chandra position was improved and the
X-ray  error box   reduced  to a  circle   of $1.6^{\prime \prime}$  radius
\cite{Garmire00}.  Optical  observations  obtained with the  1.54-m Danish
Telescope (1.54D) on 2000  Feb.  11.03-11.08 UT  ($\sim$ 16 hours after the
burst) revealed an object coincident with the X-ray error box determined by
Chandra \cite{Gorosabel00a}.   These  optical observations are  part of the
results reported in this paper.  Deep radio follow-up observations starting
14.8  hours after  the gamma-ray  event, did  not  find  any radio emission
associated to the afterglow above 55 microJy \cite{Berger00,McConnell00}.

\section{Observations}

We obtained optical observations with the  2.56-m Nordic Optical Telescope
(NOT) and the  1.54D starting 12.4   hours and 16   hours after the  burst,
respectively.  The NOT observations were carried  out with HIRAC and at the
1.54D with the DFOSC instrument.  Table 1 displays the observing log.

\begin{table}[H]
\caption{List of the observations obtained with NOT and 1.54D Telescopes.}
\begin{center}
\begin{tabular}{lclcc}
\hline
 Date 2000(UT)  & Filter & Exposure Time & Seeing & Telescope \\
\hline
Feb.     10.88 - 10.90  &  R  &  3x300      & 1.2 &  NOT \\
Feb.     11.03 - 11.08  &  R  &  10x300     & 1.6 & 1.54D\\
Feb.     14.02 - 14.03  &  R  &  600        & 1.9 & 1.54D\\
May       5.42 -  5.44  &  R  &  2x600      & 2.3 & 1.54D\\
Aug.     22.29 - 22.41  &  R  &  7x900      & 2.2 & 1.54D\\
Aug.     23.23 - 23.29  &  R  &  5x900      & 2.3 & 1.54D\\
Aug.     24.23 - 24.30  &  R  &  4x900      & 3.0 & 1.54D\\
Aug.     26.29 - 26.43  &  V  &  9x900      & 1.5 & 1.54D\\
Aug.     27.21 - 27.24  &  I  &  2x900      & 1.4 & 1.54D\\
Aug.     28.21 - 28.24  &  I  &  2x1200     & 1.4 & 1.54D\\
Aug.     29.21 - 29.30  &  I  &  7x1200     & 1.1 & 1.54D\\
Aug.     30.22 - 30.24  &  I  &  1200       & 1.1 & 1.54D\\
Aug.     31.21 - 31.24  &  B  &  2x1200     & 1.5 & 1.54D\\
\hline

\hline
\end{tabular}
\end{center}
\end{table}

\vspace{-0.8cm}

\begin{center}
\begin{figure}[H] 
 \caption{\label{chandra} The figure shows a blow up of the co-added R-band
   image taken on Aug 22.29-- 24.30 UT.  The plot shows the position of the
   improved Chandra X-ray circle \cite{Garmire00} and the optical candidate
   \cite{Gorosabel00b}.}                                                    
 \resizebox{\hsize}{!}{\includegraphics[angle=90]{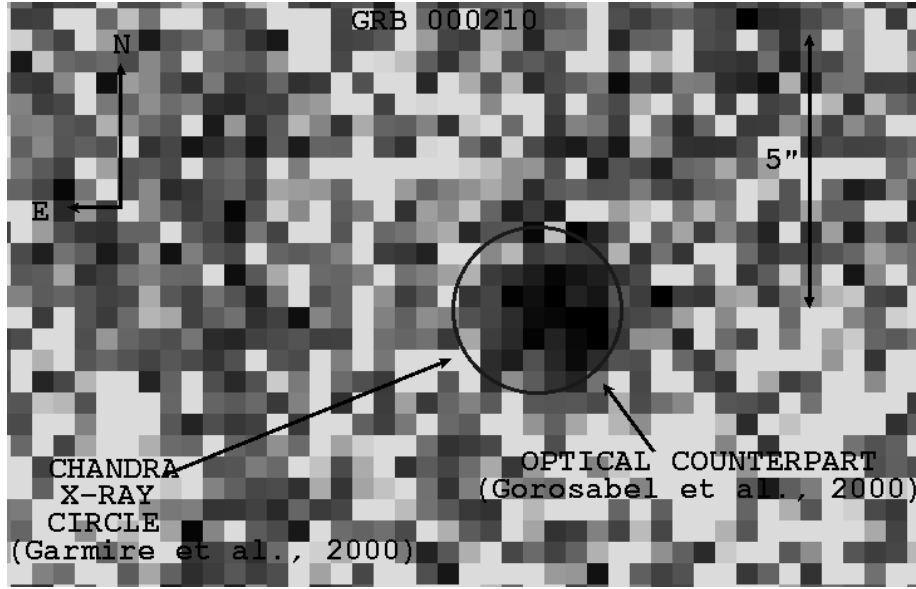}}
\end{figure}
\end{center}

The observations  carried  out in  August  improved  the  previous position
reported in February \cite{Gorosabel00a}, yielding $\alpha_{J2000} = 01^{h}
59^{m}  15.60^{s}$, $\delta_{J2000}  = -40^{\circ} 39^{\prime} 32.8^{\prime
  \prime}$    with   an     uncertainty of    ${\pm}    1^{\prime  \prime}$
\cite{Gorosabel00b}.  This position is fully  consistent with the  improved
Chandra X-ray circle position \cite{Garmire00} (see Fig \ref{chandra}). The
first images taken at  NOT were not deep  enough  to detect the object.   A
comparison of the co-added R-band image taken on Aug  22.29-- 24.30 UT with
the  R-band image  taken   16 hours  after  the   burst gives  a  magnitude
difference of 0.03 $\pm$ 0.30  mag. Therefore, the object remains  constant
in brightness within the photometric errors.  We derive a  magnitude of R =
23.5 $\pm$ 0.2 for the object.

\section{Conclusion}

Although the  object  is very  faint in  our  images its appearance  is not
stellar.   The  photo-profiles shows  an  object slightly  elongated in the
North-East  direction   with an   angular extension  of  $\sim$1.5$^{\prime
  \prime}$.  The  angular size in the  orthogonal direction (North-West) is
limited  by   the seeing (1.1$^{\prime  \prime}$  in  our best images). The
object did not change  in brightness since the  first detection carried out
just 16 hours  after the burst.   GRB 000210 appears to  be one of the best
candidates  to  study the problem  of the  ``dark   burst'', resembling GRB
970828.   It  would be   extremely  important to   perform deep  optical/IR
observations  aimed  to   determine  the   redshift,   the spectral  energy
distribution and the morphology of this enigmatic object.

\section*{Acknowledgments}
Javier  Gorosabel acknowledges the   receipt of a Marie Curie Research  Grant
from the European Commission.

%

\end{document}